\documentclass[10pt,a4paper]{article}
\usepackage[dvips]{graphicx}
\usepackage{latexsym}
\usepackage{amsmath,amssymb}

\title{Inflationary Parameters in Renormalization Group Improved $\phi^4$ Theory}
\author{\large Tomohiro Inagaki${}^1$, Ryota Nakanishi${}^2$ and Sergei D. Odintsov${}^{3,4,5}$,\\[2mm]
\normalsize
${}^1$Information Media Center, Hiroshima University, Higashi-Hiroshima, 739-8521, Japan,\\\normalsize
${}^2$Department of Physics, Hiroshima University, Higashi-Hiroshima, 739-8526, Japan,\\\normalsize
${}^3$Instituci\'{o} Catalana de Recerca i Estudis Avan\c{c}ats (ICREA), Barcelona, Spain,\\\normalsize
${}^4$Consejo Superior de Investigaciones Cient\'{i}ficas, ICE/CSIC-IEEC, Campus UAB, \\\normalsize
Torre C5-Parell-2a pl, E-08193 Bellaterra (Barcelona), Spain,\\\normalsize
${}^5$Tomsk State Pedagogical University, 634061 Tomsk and National Research Tomsk \\\normalsize
State University, 634050 Tomsk, Russia
}
\usepackage[top=30truemm,bottom=30truemm,left=25truemm,right=25truemm]{geometry}

\makeatletter
\@addtoreset{equation}{section}
\makeatother
\usepackage{braket}
\usepackage{setspace}
\begin{document}
 \maketitle

 \begin{abstract}
Inflation models can be examined by the cosmological observations, WMAP, Planck, BICEP2 and so on.
These observations directly constrain the spectral index, $n_s$, and the tensor-to-scalar ratio, $r$.
Besides, from a theoretical point of view, it has been shown that any inflation models asymptote a universal attractor in $(n_s,r)$ plane for a larger scalar-gravity coupling.
In this work we consider a simple chaotic inflation model with a scalar quartic and a scalar-curvature interactions.
The quantum corrections are introduced for these interactions through the renormalization group.
The inflationary parameters, $n_s$ and $r$, are numerically calculated with shifting the bare scalar-gravity coupling $\xi_0$, the quartic scalar bare coupling $\lambda_0$, the renormalization scale $\mu$ and the e-folding number $N$.
The Planck data is consistent with the $\phi^4$ theory with a finite scalar-curvature coupling.
It is found that the RG running induces a non-universal contribution for $n_s$ and $r$. 
It can increase the tensor-to-scalar ratio, $r$, which is consistent with Planck data so that it may approach the BICEP2 data for a small renormalization scale.
 \end{abstract}

 \section{Introduction}
Direct constraints for the inflationary parameters are obtained by observing the anisotropy of the cosmic microwave background.
Much attention have been given to the recent cosmological observations, WMAP \cite{Hinshaw:2012aka}, Planck \cite{Ade:2013uln}, BICEP2 \cite{Ade:2014xna} and so on. Many researchers use these observational results to inspect some high energy physics models in the inflationary era. Some useful indicators are the spectral index, $n_s$, and tensor-to-scalar ratio, $r$.

During the inflationary era it is not adequate to neglect the effect of the scalar-gravity coupling $\xi_0$.
The observations by Planck and WMAP support the chaotic inflation model with a scalar four-point and a strong scalar-curvature interactions \cite{Linde:1983gd}. Furthermore, a quantum field theory in curved spacetime predicts the non-zero value of parameter, $\xi_0$, in order to get multiplicative renormalizability (for a general introduction, see \cite{Buchbinder}).
The behavior of the renormalization-group effective coupling, $\xi$, depends on the theory under investigation (asymptotically-free, finite, etc) and may directly influence the inflaton dynamics (see Refs.\cite{Buchbinder} and \cite{Muta:1991mw}).
On the other hand, BICEP2 favors a scalar four-point and the minimal scalar-curvature interactions. Above results are independent of the scalar quartic coupling, $\lambda_0$, without any quantum corrections.

It is shown that any inflationary model containing a concrete non-minimal scalar-curvature coupling asymptotes a universal attractor
for a strong scalar-curvature coupling \cite{Kallosh:2013tua}. A quantum correction may change these  consequences.
The radiative corrections modify the inflaton potential. The effective potential is calculated in a weakly curved spacetime, de Sitter, anti-de Sitter spacetime and so on in the context of the  curvature induced phase transition (for a review, see Refs.\cite{Buchbinder} and \cite{Inagaki:1997kz}).
The renormalization-group (RG) improved effective potential has been investigated up to the one-loop level in Refs. \cite{Elizalde:1993qh, Buchel:2004df, DeSimone:2008ei, Lee:2013nv, Barenboim:2013wra, Okada:2013vxa, Oda:2014rpa, Herranen:2014cua, Elizalde:2014}. The two-loop calculation is given in Ref.\cite{Hamada:2014xka}. The spectral index and the tensor-to-scalar ratio have been calculated with the logarithmic corrections for the scalar self-coupling. An inflationary model which is consistent with the Planck and BICEP2 results is evaluated in Ref. \cite{Bamba:2014daa}.

As far as we know, a little number of works have been reported in the study of the primordial fluctuations with the running scalar-curvature coupling under the inflationary expanding universe. Therefore we launched our plan to make a systematic study of the inflationary parameters with the running scalar-curvature and scalar quartic couplings in a $\phi^4$ theory. In Sec.2 we introduce the leading logarithmic corrections in the $\phi^4$ theory. By generalizing the Coleman-Weinberg approach, we obtain the RG improved effective potential with the leading logarithmic corrections for the scalar-curvature and the scalar quartic couplings \cite{Elizalde:1993ee}. The explicit expressions for the spectral index and tensor-to-scalar ratio are obtained under the RG improved effective potential in Sec.3.
We numerically evaluate the inflationary parameters in Sec.4. Some typical behaviors are shown for the spectral index and tensor-to-scalar ratio. We briefly discuss the consistency with the constraints from the cosmological observations. Finally we give some concluding remarks. 

 \section{RG-improved effective Lagrangian in Einstein frame}
We consider a simple chaotic inflation model. 
A scalar four-point interaction is employed to induce inflationary expansion of the universe. 
We start with the Lagrangian consisting of the Einstein-Hilbert term and the scalar field with a quartic potential \cite{Kallosh:2013maa},
\begin{eqnarray}
\mathcal{L}^{(J)} &=& \sqrt{-g}\left(\frac{1}{2}R + \frac{1}{2}\xi_0 R\phi^2 - \frac{1}{2}g^{\mu\nu} (\partial_{\mu} \phi)( \partial_{\nu}\phi) - V^{(J)} \right), \label{lag:J} \\
V^{(J)} &=& \frac{\lambda_0}{4!}\phi^4 ,
\end{eqnarray}
where the superscript $(J)$ denotes the Jordan frame, $g$ is the determinant of the metric tensor, $g_{\mu\nu}$, $\xi_0$ and $\lambda_0$ represent the bare scalar-curvature and scalar quartic couplings, respectively. In the Jordan frame the scalar-curvature coupling is explicitly appears.

In the quantum field theory the classical Lagrangian is modified by the quantum corrections.
Following the generalized Coleman-Weinberg approach developed in Ref.\cite{Elizalde:1993ee}, see also \cite{Buchbinder}, the effective Lagrangian up to the leading log order is obtained by the replacements,
\begin{eqnarray}
\xi_0\ \to\ \ \xi &=& \frac{1}{6} + \left(\xi_0 - \frac{1}{6} \right)\left ( 1 - \frac{3\lambda_0}{32\pi^2}\ln{\frac{\phi^2}{\mu^2}}\right)^{-1/3},
\label{replace1}
\\
\lambda_0\ \to\ \ \lambda &=& \frac{\lambda_0}{1 - \cfrac{3\lambda_0}{32\pi^2}\ln{\cfrac{\phi^2}{\mu^2}}},
\label{replace2}
\end{eqnarray}
where $\xi$ and $\lambda$ denote the running couplings and $\mu$ represents the renormalization scale.
It is easy to find that the scalar-curvature coupling, $\xi$, has a conformal fixed point at $\xi=1/6$.

Applying the replacements (\ref{replace1}) and (\ref{replace2}) for the Lagrangian (\ref{lag:J}), we obtain the one-loop RG improved effective Lagrangian, $\mathcal{L}_{eff}^{(J)}$,
\begin{eqnarray}
\mathcal{L}_{eff}^{(J)} &=& \sqrt{-g}\left(\frac{1}{2}R + \frac{1}{2}\xi R\phi^2 - \frac{1}{2}g^{\mu\nu} (\partial_{\mu} \phi)(\partial_{\nu} \phi) - V_{eff}^{(J)} \right),\\
V_{eff}^{(J)} &=& \frac{\lambda}{4!}\phi^4 .
\label{lag:eff:J}
\end{eqnarray}
In practical calculations it is more convenient to evaluate the theory in the Einstein frame where the scalar-curvature coupling disappears. We consider the Weyl transformation,
\begin{equation}
\tilde{g}^{\mu\nu} = \Omega^{-2} g^{\mu\nu}.
\label{Weyl}
\end{equation}
By this Weyl transformation the one-loop RG improved effective Lagrangian becomes
\begin{eqnarray}
\mathcal{L}^{(J)}_{eff} &\rightarrow& \Omega^{-2}\sqrt{-\tilde{g}}\biggl[\frac{1}{2}\tilde{R} + \frac{1}{2}\xi\tilde{R}\phi^2 - \frac{1}{2}\tilde{g}^{\mu\nu}(\partial_{\mu} \phi) (\partial_{\nu} \phi) - \Omega^{-2}V^{(J)}_{eff} \nonumber \\
&\ &\ \ \ \ \ \ \ \ \ \ \ \ \ \ \ \ \ \ \ \ + 3\left[\tilde{\square}\ln{\Omega} - \tilde{g}^{\mu\nu}(\partial_{\mu}\ln{\Omega})(\partial_{\nu} \ln{\Omega})\right]\left(1 + \xi\phi^2 \right)\biggr].
\end{eqnarray}
The transformation from the Jordan frame to the Einstein frame is found by setting the Weyl factor, $\Omega$, to eliminate the scalar-curvature coupling,
\begin{equation}
\Omega^2 = 1 + \xi \phi^2.
\label{para:Weyl}
\end{equation}
The Weyl transformation modifies the kinetic term for the scalar field. The scalar field is redefined to satisfy
\begin{equation}
\frac{\partial\varphi}{\partial\phi} = \frac{\sqrt{\Omega^2 + \cfrac{3}{2} \left(\cfrac{\partial\Omega^2}{\partial\phi}\right)^2}}{\Omega^2}.
\end{equation}
Owing to this redefinition of the field the non-canonical terms vanish in the Lagrangian. Therefore we eventually reach a simple form for the canonical Lagrangian in the Einstein frame,
\begin{eqnarray}
\mathcal{L}^{(E)}_{eff} &=& \sqrt{-\tilde{g}}\left[\frac{1}{2}\tilde{R} - \frac{1}{2}\tilde{g}^{\mu\nu} \partial_{\mu} \varphi \partial_{\nu} \varphi - V^{(E)}_{eff} \right],\\
V^{(E)}_{eff} &=& \Omega^{-4}V^{(J)}_{eff},
\label{lag:E}
\end{eqnarray}
where the superscript $(E)$ denotes the Einstein frame. The renormalization scale dependence is included in the effective potential, $V^{(E)}_{eff}$.

 \section{Inflationary parameters}
CMB observation directly constrains the parameters of the slow-roll scenario. 
In the scenario the cosmological parameters are fixed by the form of the effective potential.
The e-folding number, $N$, and the slow-roll parameters, $\epsilon$ and $\eta$, are calculated by the following forms \cite{Kallosh:2013tua},
\begin{eqnarray}
N &=& \int_{\phi_{end}}^{\phi_N} \left(\frac{\partial\varphi}{\partial\phi}\right)^2 \frac{V^{(E)}}{\partial V^{(E)} /\partial\phi}\ d\phi, 
\label{cp1}\\
\epsilon &=& \frac{1}{2}\left(\frac{1}{V^{(E)}}\frac{\partial V^{(E)}}{\partial\phi}\frac{\partial\phi}{\partial\varphi} \right)^2 , 
\label{cp2}\\
\eta &=& \frac{1}{V^{(E)}}\left[\frac{\partial}{\partial\phi}\left(\frac{\partial V^{(E)}}{\partial\phi}\frac{\partial\phi}{\partial\varphi}\right) \right]\frac{\partial\phi}{\partial\varphi} , 
\label{cp3}
\end{eqnarray}
where the upper and the lower limits of the integral, $\phi_N$ and $\phi_{end}$, represent the field configurations when the slow-roll scenario starts and breaks. The field configuration rolls down from $\phi_N$ to $\phi_{end}$ during the inflation era.
We apply the one-loop RG improved effective potential for these formulae.
Substituting Eq.(\ref{lag:E}) with Eqs.(\ref{lag:eff:J}) and (\ref{para:Weyl}) into Eqs. (\ref{cp1}), (\ref{cp2}) and (\ref{cp3}), we obtain
\begin{eqnarray}
N &=& \int_{\phi_{end}^2}^{\phi_N^2} \frac{\Omega^2 + \cfrac{3}{2}\bigl(\Omega^{2'}\bigr)^2}{2\Omega^2\left[\left(4 + \cfrac{3\lambda}{16\pi^2} \right)\Omega^2 - 2\Omega^{2'}\phi\right]}  d\phi^2,
\label{cp4}\\
\epsilon &=& \frac{\cfrac{1}{2}\left[\left(4 + \cfrac{3\lambda}{16\pi^2} \right)\phi^{-1}\Omega^2 -2\Omega^{2'} \right]^2}{\Omega^2 + \cfrac{3}{2}\left(\Omega^{2'}\right)^2},
\label{cp5} \\
\eta &=& \left\{
\left[\left(12 + \frac{21\lambda}{16\pi^2} + \frac{9\lambda^2}{128\pi^4}\right) \phi^{-2}\Omega^4 -3\left(4 + \frac{3\lambda}{16\pi^2}\right) \phi^{-1}\Omega^2\Omega^{2'} + 4\bigl(\Omega^{2'}\bigr)^2 - 2\Omega^2\Omega^{2''} \right]\right. \nonumber \\
& & \times \left[ \Omega^2 + \frac{3}{2}\bigl(\Omega^{2'}\bigr)^2\right] \nonumber\\
& & - \ \left.\left[\left(2 + \frac{3\lambda}{32\pi^2} \right)\phi^{-1}\Omega^4 - \Omega^2\Omega^{2'} \right]\left(\Omega^{2'} + 3\Omega^{2'}\Omega^{2''}\right)\right\} \left[\Omega^2 + \frac{3}{2}\left(\Omega^{2'}\right)^2\right]^{-2},
\label{cp6}
\end{eqnarray}
where we define
\begin{eqnarray}
\Omega^{2'} &\equiv& \frac{\partial\Omega^2}{\partial\phi} 
= \left[2\xi + \frac{\lambda (\xi - 1/6)}{16\pi^2}\right] \phi, \label{Omega1} \\
\Omega^{2''} &\equiv& \frac{\partial^2\Omega^2}{\partial\phi^2} 
=  2\xi + \frac{\lambda(\lambda/4\pi^2+3)}{16\pi^2}\left(\xi - \frac{1}{6}\right).
\label{Omega2}
\end{eqnarray}

Under the slow-roll approximation the spectral index, $n_s$, and the tensor-to-scalar ratio, $r$, are obtained by
\begin{eqnarray}
n_s &=& 1 + 2\left.\eta\right|_{\phi=\phi_N} - 6\left. \epsilon\right|_{\phi=\phi_N}  ,
\label{ns}\\
r &=& 16\left. \epsilon\right|_{\phi=\phi_N}  .
\label{r}
\end{eqnarray}
In the slow-roll approximation we assume that the slow-roll parameters are small enough, $\eta \ll 1$ and $\epsilon \ll 1$.
We fix the field configuration, $\phi_{end}$,  when the slow-roll scenario breaks down as $\eta$ or $\epsilon$ grow up of order one. It should be noted that the field configuration, $\phi_{end}$ has only a little contribution to the e-folding number, $N$.
Thus the observational constraint for the e-folding number specifies the field configuration, $\phi_{N}$.

Taking the limit $\lambda_0 \to 0$, both the coupling constants, $\lambda$ and $\xi$, are fixed. 
At this limit we can analytically perform the integration in Eq.(\ref{cp4}).
Thus Eqs.(\ref{cp4}), (\ref{cp5}) and (\ref{cp6}) are simplified to
\begin{eqnarray}
N &=& \frac{1}{8} \left[ (1+6\xi_0)(\phi_N^2 - \phi_{end}^2) -6\ln{\frac{1+\xi_0 \phi_N^2}{1+\xi_0 \phi_{end}^2}} \right],
\label{lim:N:nolam}\\
\epsilon &=& \frac{8}{\phi^2(1 +\xi_0 \phi^2 + 6\xi_0^2 \phi^2)} ,
\label{lim:ep:nolam}\\
\eta &=& \frac{4(3 + \xi_0 \phi^2 +12 \xi_0^2 \phi^2 - 2\xi_0^2 \phi^4 - 12\xi_0^3 \phi^4)}{\phi^2(1+\xi_0 \phi^2 + 6\xi_0^2 \phi^2)^2}.
\label{lim:et:nolam}
\end{eqnarray}
These expressions coincide with the ones obtained in Ref.\cite{Kallosh:2013maa}.
It should be noticed that the inflationary parameters (\ref{lim:N:nolam}), (\ref{lim:ep:nolam}) and (\ref{lim:et:nolam}) are independent of the coupling constant, $\lambda_0$. The parameters, $n_s$ and $r$, are obtained by fixing the scalar-curvature coupling $\xi_0$. 
It is easy to find that 
\begin{equation}
\left.\epsilon\right|_{\phi=\phi_N} \rightarrow \frac{3}{4N^2},\, \, \,
\left.\eta\right|_{\phi=\phi_N} \rightarrow -\frac{1}{N} ,
\label{lim:strong1}
\end{equation}
at the large $\xi_0$ limit with $\phi_N\gg \phi_{end}$. Thus the parameters, $n_s$ and $r$, approach to the universal attractor,
\begin{equation}
r \rightarrow \frac{12}{N^2},\, \, \,
n_s \rightarrow 1-\frac{2}{N} ,
\label{lim:strong}
\end{equation}
with the growth of $\xi_0$ for $\lambda_0 = 0$. These equations reproduce the results in Ref.\cite{Kallosh:2013tua}.

 \section{Numerical results}
In the previous section we calculate the inflationary parameters in the slow-roll approximation.
Here we vary four parameters, the bare scalar-curvature coupling, $\xi_0$, the renormalization scale, $\mu^2$, the bare quartic scalar coupling, $\lambda_0$, and the e-folding number, $N$ and evaluate the inflationary parameters, $n_s$ and $r$, in the following procedure.
We set the field configuration, $\phi_{end}$, by
\begin{equation}
\left.\epsilon\right|_{\phi=\phi_{end}}=1 .
\end{equation}
Numerically performing the integration in Eq.(\ref{cp4}) as the e-folding number, $N$, fixes, we obtain the initial field configuration, $\phi_{N}$. As is shown in Fig.1, the configuration, $\phi_N$, is sensitive to the scalar curvature coupling, $\xi_0$, and much larger than $\phi_{end}$. It should be noticed that all the mass scale are normalized by the Planck scale. The slow-roll parameters, $\left.\epsilon\right|_{\phi=\phi_N}$ and $\left.\eta\right|_{\phi=\phi_N}$,  are calculated from Eqs.(\ref{cp5}) and (\ref{cp6}). If these parameters are small enough, the slow-roll approximation is applicable and the inflationary parameters are obtained by Eqs.(\ref{ns}) and (\ref{r}).

\begin{figure}
  \begin{center}
   \includegraphics[width=72mm]{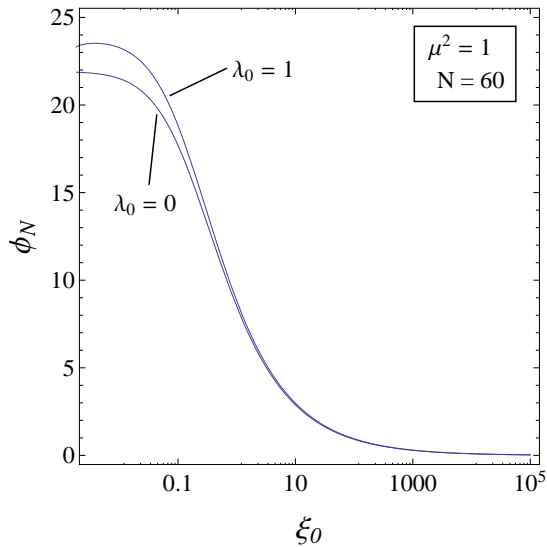}
  \end{center}
  \caption{Behavior of $\phi_N$ as a function of $\xi_0$ for $N=60$ and $\mu^2=1$. }
\end{figure}

\begin{figure}
 \begin{minipage}{0.48\hsize}
  \begin{center}
   \includegraphics[width=71mm]{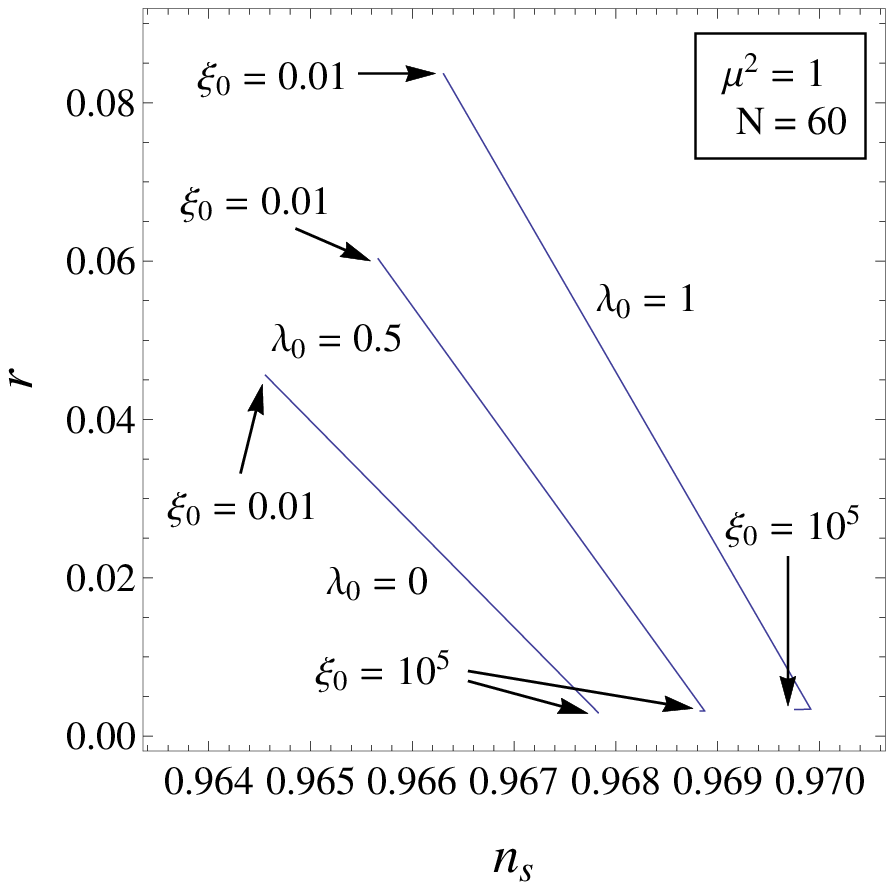}
  \end{center}
  \caption{Behavior of $n_s$ and $r$ as a function of the bare coupling $\xi_0$ in the interval from $0.01$ to $10^5$ for $N=60$, $\mu^2=1$ and $\lambda_0 = 0,0.5,1$.}
 \end{minipage}
 \hspace{0.04\hsize}
 \begin{minipage}{0.48\hsize}
  \begin{center}
   \includegraphics[width=69mm]{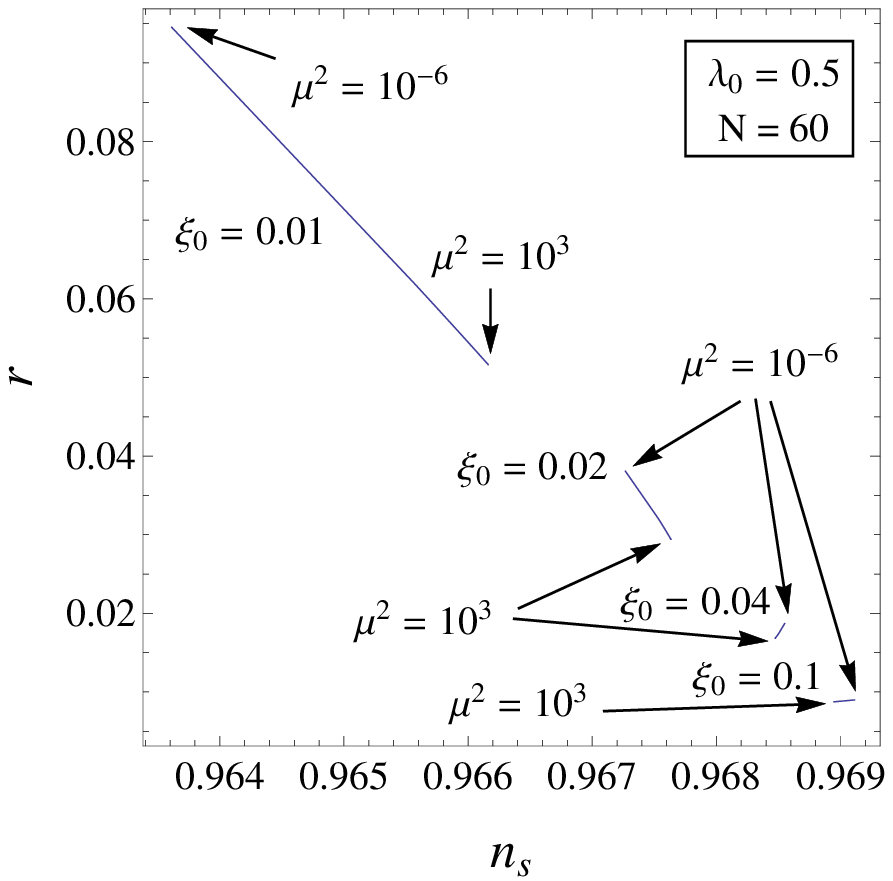}
  \end{center}
    \caption{Behavior of $n_s$ and $r$ as a function of the renormalization scale $\mu^2$ in the interval from $10^{-6}$ to $10^3$ for $N=60$, $\lambda_0=0.5$ and $\xi_0=0.01,0.02,0.04,0.1$.}
 \end{minipage}
\end{figure}

In Fig. 2 we illustrate the $\xi_0$-dependence of $n_s$ and $r$. In Ref.\cite{Kallosh:2013tua} a universal attractor is found for a wide class of chaotic inflation models. In these models the inflationary parameters, $n_s$ and $r$, reach the attractor for a strong scalar-curvature coupling. The results in Fig. 2 shows that the quantum corrections give non-universal influences on $n_s$ and $r$ until $\xi_0=10^5$.

We present the renormalization scale dependence in Fig. 3. The $\mu$-dependence has some similar properties with the $\xi_0$-dependence. As either $\xi_0$ or $\mu^2$ increases, the tensor-to-scalar ratio $r$ decreases. Both trajectories almost overlap as $\xi_0$ or $\mu$ varies. For a stronger scalar-curvature coupling $\xi_0$ the renormalization scale dependence is suppressed. The attractor vanishes or moves to a larger $\xi_0$ by the leading logarithmic corrections, but there is a possibility to define a renormalization scale independent results with the growth of the scalar-curvature coupling.

\begin{figure}
 \begin{minipage}{0.48\hsize}
  \begin{center}
   \includegraphics[width=72mm]{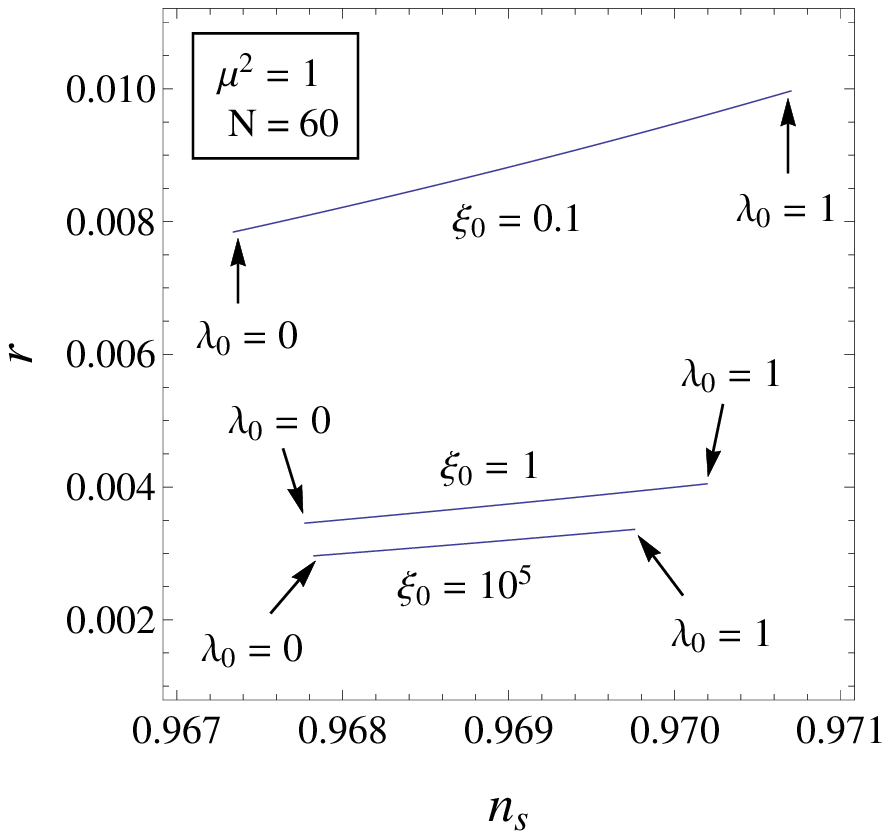}
  \end{center}
    \caption{Behavior of $n_s$ and $r$ as a function of the bare coupling $\lambda_0$ in the interval from $0$ to $1$ for $N=60$, $\mu^2=1$ and $\xi_0 = 0.1,1, 10^5$.}
 \end{minipage}
  \hspace{0.04\hsize}
 \begin{minipage}{0.48\hsize}
  \begin{center}
   \includegraphics[width=70mm]{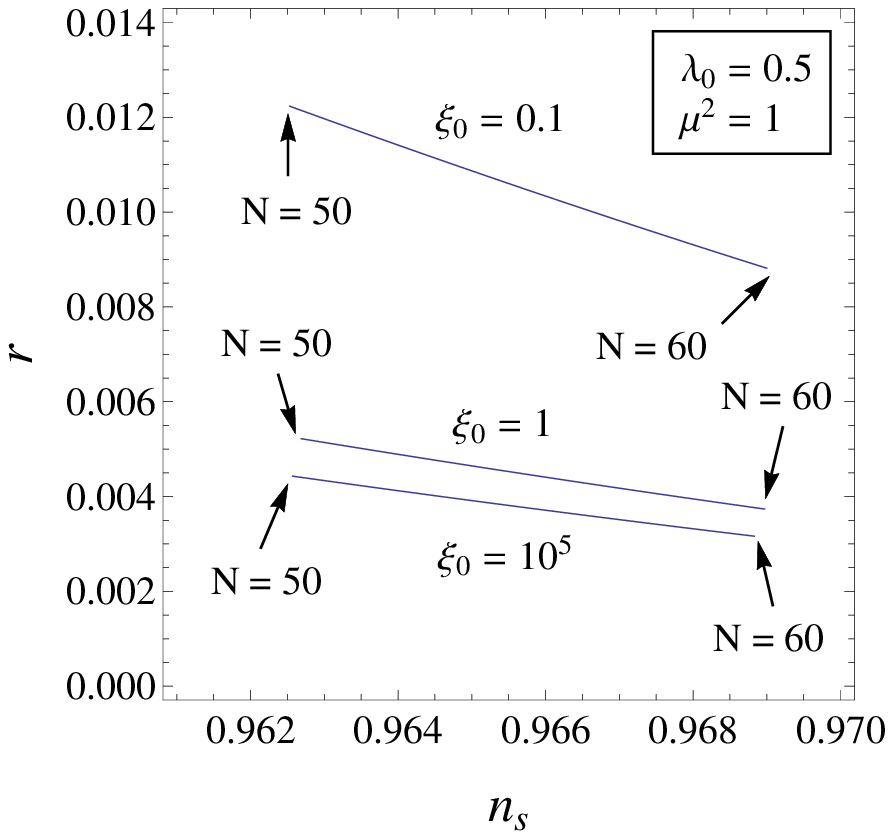}
  \end{center}
      \caption{Behavior of $n_s$ and $r$ as a function of the e-folding number $N$ in the interval from $50$ to $60$ for $\lambda_0=0.5$, $\mu^2=1$ and $\xi_0 = 0.1,1, 10^5$.}
 \end{minipage}
\end{figure}

In Fig. 4 we represent the $\lambda_0$-dependence of $n_s$ and $r$. The running couplings are perturbatively calculated at the leading order of $\lambda_0$. Thus we change the scalar quartic coupling in the interval from $0$ to $1$ in order to keep the validity of the perturbative expansion. It is observed that both $n_s$ and $r$ increases as the quartic scalar coupling, $\lambda_0$, grows. If we consider a smaller renormalization scale, $\mu^2<\phi_N^2$, we observe different behavior. The spectrum index, $n_s$, decreases as the quartic scalar coupling, $\lambda_0$, grows. It describes the non-universal influences on $n_s$ and $r$ until $\xi_0 = 10^5$. The contribution to the tensor-to-scalar ratio is suppressed as the scalar-curvature coupling, $\xi_0$, increases.

The dependence on the e-folding number, $N$, is shown in Fig. 5. In this figure the spectral index $n_s$ increases about $0.6\%$ and the tensor-to-scalar ratio $r$ decreases $12-30\%$ as $N$ increases. We see a similar behavior with the one in Ref.\cite{Kallosh:2013tua}. The RG running has no significant influence on the  $N$-dependence.

\begin{figure}
  \begin{center}
   \includegraphics[width=72mm]{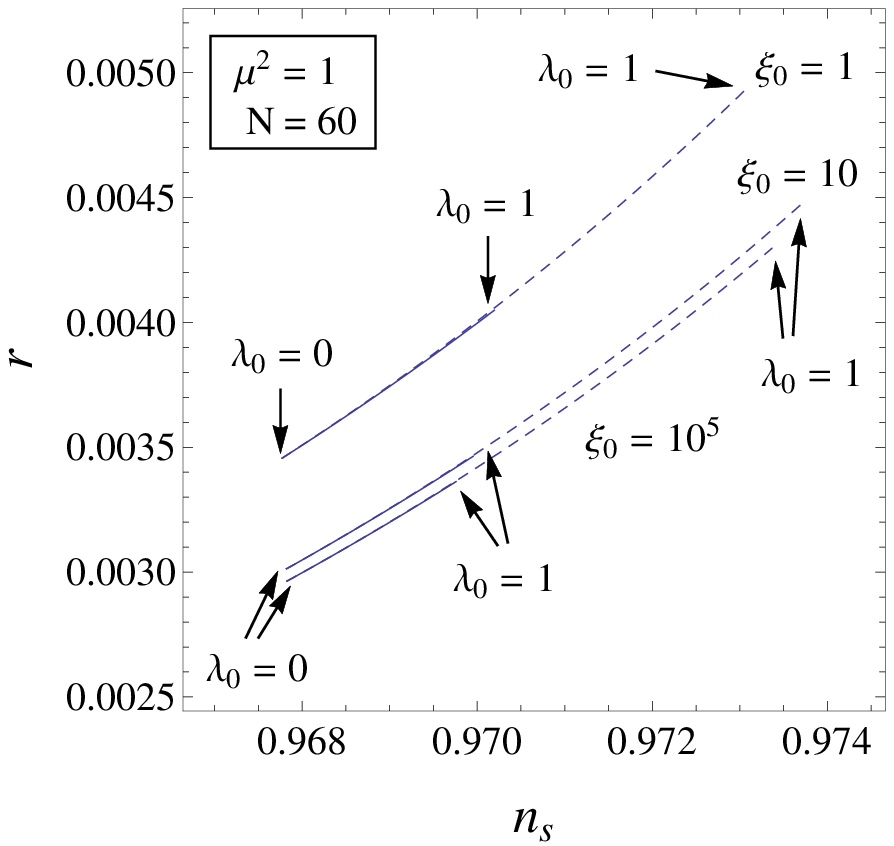}
  \end{center}
  \caption{Behavior of $n_s$ and $r$ as a function of $\lambda_0$ for $\mu^2=1$, $N=60$ and $\xi_0 = 1, 10, 10^5$. We include the RG-running for both scalar quartic coupling, $\lambda$, and  scalar-curvature coupling, $\xi$ and plot the solid lines. The dashed lines are drawn with eliminating the RG-running for the scalar-curvature coupling, $\xi$.}
\end{figure}

We want to see the contribution from running of the scalar-curvature coupling. To distinguish this contribution we set $\xi(\mu)=\xi_0$. In this case the scalar-curvature coupling is fixed at  $\xi_0$. Then Eqs.(\ref{Omega1}) and (\ref{Omega2}) simplify to
\begin{equation}
\Omega^{2'} = 2\xi_0 \phi,  \, \,
\Omega^{2''} =  2\xi_0 .
\label{Omega:fix:xi}
\end{equation}
Evaluating Eqs.(\ref{cp4}), (\ref{cp5}) and (\ref{cp6}) with Eq.(\ref{Omega:fix:xi}), we find $n_s$ and $r$ for running only the scalar quartic coupling, $\lambda$. We show the results in Fig. 6. The dashed lines represent the behavior as $\xi$ fixes at $\xi=\xi_0$. For $\lambda_0=0$ both the coupling constants fix at $\lambda=\lambda_0=0$ and $\xi=\xi_0$. At the fixed point the running effect disappears. Thus each dashed lines reaches the same point with the corresponding solid lines at $\lambda_0=0$. It is also observed that the $\lambda_0$-dependence is suppressed if we introduce the running for $\xi$.
 
Under the slow-roll scenario the inflationary parameters, $n_s$ and $r$, are estimated from the Planck in combination
with WMAP 9-year large angular scale polarization data \cite{Ade:2013uln}. It is possible to reproduce the central value for the spectrum index, $n_s = 0.9653$ by tuning the scalar-curvature coupling, $\xi_0=0.014$, without RG running. The RG-running can increase the tensor-scalar ratio, $r$. As is shown in Tab.1, the ratio, $r$, increases about $70\%$ for the parameter set, $\lambda_0=0.5, \mu^2=10^{-6}$, by the RG-running effect and approach BICEP2 constraint for a small renormalization scale.

\begin{table} 
\begin{center}
\begin{tabular}{c|c|c|c|c}
 & Planck+WP+lensing & $\lambda_0=0$ & $\lambda_0=0.5, \mu^2=1$ & $\lambda_0=0.5, \mu^2=10^{-6}$ \\\hline
 $n_s$ & $0.9653\pm0.0069$ & 0.9653 & 0.9667 & 0.9659\\
 $r$ & $<0.13$ & 0.035 & 0.044 & 0.059\\
 \end{tabular}
 \caption{Planck constraint for $n_s$ and $r$ \cite{Ade:2013uln} and our results for $\xi_0=0.014$ and $N=60$.}
 \end{center}
 \end{table}
 
 \section{Conclusions}
We have studied the quantum effect on the inflationary parameters, $n_s$ and $r$, in a simple $\phi^4$ theory with non-minimal scalar-curvature coupling. The effective Lagrangian is obtained up to the leading log order by the generalized Coleman-Weinberg approach. A quantum effect is introduced through the RG-running of the scalar quartic coupling, $\lambda$, and the scalar-curvature coupling, $\xi$. 
Assuming the slow-roll scenario, we systematically evaluate the typical behavior for $n_s$ and $r$ as a function of the bare couplings, the renormalization scale or the e-folding number.

It is found that the quantum corrections have a non-universal contribution for $n_s$ and $r$. The inflationary parameters, $n_s$ and $r$, do not reach the attractor (\ref{lim:strong}) until $\xi_0=10^5$. However, the $\mu^2$- and $\lambda_0$-dependences for $n_s$ and $r$ suppressed as the scalar-curvature coupling, $\xi_0$, increases. Thus the inflationary parameters, $n_s$ and $r$, at least approximately reach the attractor at the strong scalar-curvature coupling limit. 

Our model is a simple toy model but we discuss the consistency with the constraints from the cosmological observations. It is known that the Planck constraint is consistent with the $\phi^4$ theory with a finite scalar-curvature coupling \cite{Ade:2013uln}. The RG-running introduces $\lambda_0$- and $\mu$-dependences for $n_s$ and $r$. It can increase the tensor-scalar ratio, $r$, to the value consistent with the BICEP2 data.

The present work is restricted to the analysis of the leading logarithmic quantum corrections for the inflationary parameters in the simplest model. We are interested to apply our result to realistic models. For this purpose we should include the contribution from the weak gauge and the Yukawa type scalar-fermion interactions. Using the results in Ref.\cite{Odintsov:1993rt}, one can study the model up to the two-loop level and study the contribution from a non-trivial fixed point of the RG. The vacuum stability of the effective potential should be also considered  to find a more realistic constraint \cite{Kobakhidze:2013tn}.

\section*{Acknowledgements}
The authors would like to thank K. Yamamoto for fruitful discussions.
TI is supported by JSPS KAKENHI Grant Number 26400250. 
SDO is supported in part by MINECO (Spain), FIS2010-15640 and by MES project TSPU-139 (Russia).

\end{document}